# Annihilation of Relativistic Positrons in Single Crystal with production of One Photon

Kalashnikov N.P., Mazur E.A., Olczak A.S.

National Research Nuclear University "MEPhI" (Moscow Engineering Physics Institute), Moscow, Russia

## 1. Introduction

Normally in the process of electron-positron annihilation when both leptons were free two photons are produced, as the annihilation process with production of one photon is forbidden by the conservation laws (for example [1, 2]). But in case, when one or both leptons are not free (electron belongs to atom, for example) this process may become possible [1, 3]. Another possibility for this process opens in the case, when one of the leptons (e.g. high energy relativistic positron) is moving in single crystal in channeling or quasi-channeling mode, when its motion is either bound or sufficiently uneven [4,5].

## 2. Kinematics of the single photon annihilation of relativistic positron

If the relativistic positron propagates along the axis OZ with energy $E_+ \gg m_e c^2$, the new produced single photon must move also along the axis OZ, what is evidently forbidden by momentum and energy conservation laws: (energy: $E_+ + m_e c^2 = \hbar\omega$; momentum: $P_+ = \hbar\omega /c$. Consequence: $P_+ c < E_+ < E_+ + m_e c^2 = \hbar\omega \Rightarrow \hbar\omega$ cannot be equal to $P_+ c$).

Momentum conservation law however can be satisfied in projection upon axis OZ, if the newly produced photon propagates under small angle θ to the axis OZ:

$$P_+ c = \hbar\omega \cos\theta \approx \hbar\omega (1-\theta^2/2) < \hbar\omega \qquad (1)$$

In this case transverse momentum $P_x$ is not conserved, what is forbidden in vacuum. However, it may be compensated by an interaction with the crystal lattice, for example if the positron moves in channeling or quasi-channeling mode, when its motion is either bound between neighboring crystal planes (channeling mode) or hovering over crystal planes (quasi-channeling mode).

Let us estimate from conservation laws the possible angles θ at which the photon may propagate in case of single-photon annihilation. When $E_+ \gg m_e c^2$, we may expect that θ≪1. Energy conservation law looks like:

$$E_+ + m_e c^2 = \hbar\omega \qquad (2)$$

For the longitudinal momentum $P_z$ the conservation law looks like:

$$P_z c = \hbar\omega \cos\theta \approx \hbar\omega (1-\theta^2/2) \qquad (3)$$

Energy depends on momentum in relativistic case as:

$$E_+ = (P_+^2 c^2 + m_e^2 c^4)^{1/2} = (P_z^2 c^2 + 2E_+ E_x + m_e^2 c^4)^{1/2}, \qquad (4)$$

where $E_x$ - is the transverse energy of the positron, moving in averaged continuous channeling potential [4,5]. Comparing (2) and (4) the result for relativistic positron energies will have the following form:

$$P_zc(1 + E_x/E_+ + m_e^2c^4/2E_+^2) + m_ec^2 = \hbar\omega \tag{5}$$

Substituting $P_zc$ (3) we obtain:

$$\hbar\omega - \hbar\omega\theta^2/2 + E_x + m_e^2c^4/2E_+ + m_ec^2 = \hbar\omega \tag{6}$$

Taking into account, that $E_x << m_ec^2$ [4,5] and $m_e^2c^4/2E_+ << m_ec^2$ we get the result:

$$\hbar\omega\theta^2 \sim m_ec^2 => \theta \sim (m_ec^2/\hbar\omega)^{1/2} \sim (m_ec^2/E_+)^{1/2} >> m_ec^2/E_+ \tag{7}$$

### 3. Some considerations for planar channeling case

Channeling phenomena occurs when a charged particle enters a single crystal at a small angle with respect to the crystallographic axis or plane (for example, [4,5,6]). The entrance angle θ must be smaller or comparable with the Lindhard angle [4,5,6],

$$\theta_L \sim (2U_0/E_+)^{1/2} << 1, \tag{8}$$

where $U_0 \sim Ze^2R_{at}/d^2$ is the effective depth of the continuous channeling potential (for planar channeling), $R_{at}$ is the atom radius, $d$ is the lattice constant ($U_0 \sim$ 20-50 eV for planar channeling in most of crystals [4]). The channeling lepton energy $E_+$ must be much higher than the rest energy of the lepton $E_+ >> m_ec^2 =$ 511 keV.

To produce a photon with transverse energy $\sim \hbar\omega\theta$ the initial positron should have the comparable value of transverse momentum and energy.

$$(2E_+E_x)^{1/2} \sim \hbar\omega\theta \sim (m_ec^2E_+)^{1/2} \tag{9}$$

Actually the transverse energy of positron in planar channeling potential is comparable to its height $E_+ \sim U_0 << m_ec^2$. That means that in channeling mode the positron has much less transverse energy to satisfy the necessary condition (9), which strongly reduces the possibility to produce single photon in annihilation with free electron, though this process is described by simple one vertex diagram Fig.1

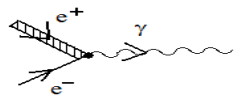

Fig.1. *Single-vertex Feynman diagram for single-photon annihilation of channeled positron on free electron.*

However, the planar channeling motion provides yet one more possibility to increase the cross section of single-photon annihilation. We will discuss it in the next section.

### 3. Specifics of positron motion in planar channeling mode

Motion of channeling particles was thoroughly studied in different aspects by many researchers in 1970-80-ies (see for example [4,5,6]). The wave function of the positively charged particles which move in the planar channel can be presented as

$$\psi^{(+)}(\vec{r}) = \Sigma_k \exp(ip_z z) u_k(x), \qquad (10)$$

For the transversal component of the positron wave function $u_k(x)$ we may use the one-dimensional Schrödinger equation with relativistic mass $E_+$ [4, 5] and periodical border conditions

$$u''_k(x) + 2E_+(\varepsilon_k - U(x))u_k(x) = 0 \qquad (11)$$

$$u_k(x+d) = u_k(x) \qquad (12)$$

where $\varepsilon_k$ – are the allowed transversal energies of the positron motion in the planar channel with averaged potential $U(x)$, numbered by index $k = 1,2,3…$ The transverse energies of a particle in periodical planar channeling potential form a structure of energy zones, which for the deep under-barrier energies can be considered as discrete levels. Calculation of zone structure with realistic continuous potential by averaging the potentials of atoms, constituting the crystal plane [4, 5,6] is a task for numerical calculations. Fig.2(a) shows numerical simulation of such zones for 25 MeV positrons in (110) planar channel in Si. Population of zones (Fig.2(b)) is depending on the incidence angle of the particle relative to the crystal plane. Fig.3 shows the numerical simulation of the averaged wave function square module (probability to find a positron in this or that part of a channel) for different zones. For high energy levels (where quasi-classical approach is applicable) the probability density to find a particle near coordinate x is proportional to $\sim (\varepsilon_k - U(x))^{-1/2}$. That means that under-barrier particles are moving mostly close to the turning points, where $\varepsilon_k = U(x)$, and for over-barrier particles probability density is maximal over the barrier. As a consequence - positrons with close to the barrier energy levels (both over and under potential barrier) are hovering over the atomic planes, where the electron density is $\sim d/R \sim 10$ times higher, than average. For such particles cross sections for all the processes of interactions with atomic electrons, including the single-photon annihilation process on atomic electrons, shall be multiplied by approximately the same factor $\sim d/R$. To achieve this it is necessary to populate by positrons mostly close to the barrier levels, what is possible when the entrance angle of positron beam in relation to the atomic plane is close to the Lindhard angle [4,5].

As is known, the crystal potential $V(\vec{r})$ for crystals, such as silicon, germanium, InSb, GaAs, should have the following form if the point group of the crystal contains the inversion $V_{-\vec{G}} = V_{\vec{G}}$

,
$$V(x) = \sum_{G_x} V_{G_x} \exp(iG_x x) = V_0 + 2\sum_{G_x>0} V_{G_x} \cos(G_x x) \qquad (13)$$

Take concrete lattice potential with the diamond structure in the definite direction of channeling leading to the equidistance channeling potential planes. For a significant part of these crystals spatially inhomogeneous terms in the averaged planar crystal potential (13) can be taken zero, and the remaining terms are expressed in terms of the Fourier components $V_{4n,0,0}(n=1,2,3...)$ of the spatial lattice potential, so that

$$V(x) = \overline{V} + 2\sum_{n=1,2} V_{4n,0,0} \cos(4Gx). \qquad (14)$$

Confining ourselves to the first two terms in the sum (14), we can analytically consider the problem of channeling of the charged particle. In this case, the equation (11) is reduced to the known Mathieu equation [7] with the dimensionless coefficients

$$\frac{\partial^2 U}{\partial S^2} + (\tilde{a} - 2q\cos 2S)U(S) = 0, \qquad (15)$$

where

$$\tilde{a} = E_\perp^2 / 4G^2\hbar^2 c^2 - \frac{8EV_0}{2\hbar^2 c^2 G^2}; \quad q = \frac{EV_G}{2\hbar^2 c^2 G^2}; \quad S = 2Gx; \quad G \equiv G_{\min} = 2\pi/d; \qquad (16)$$

$d$ is the lattice constant. In the case of channeled positively charged particles $q > 0$, since $\overline{V}$ and $V_{400}$ are larger than zero. In the case of negatively charged particles $\overline{V} < 0$ as $-V_{400}$. After the shift of the origin of energy $q$ can be considered positive and for the negative particles. The parameter $q$ in both cases can be both smaller and greater than zero, depending on the value of the channeling particle (CP) transverse energy $E_\perp(P_x)$. Estimate the $q$ value in the equation (16). At $E = 28$ MeV, $\overline{V} \approx 15$ eV (i.e., the depth of the potential wall is supposed to be equal to 30 eV, $G_x = 10^{10}$ 1/m), we obtain $q \approx 11.2$. The parameter $q$ varies depending on $E_\perp(P_x)$, and can be both large and small. The solutions of (15) have a zone Bloch character. The boundaries of the energy bands are determined by the well-known in the theory of Mathieu functions [7] numbers, $a_r$ and $b_{r+1}$, where $r = 0,1,2...$, and the allowed values of the CP transverse energy $E_\perp(P_x)$ should be determined from $a_r < E_\perp(P_x) < b_{r+1}$ ($r$ is the number of the allowed zone). It is clear that the referencing of certain areas to the discrete or continuous spectrum is nominal and is determined exclusively by the CP band width. Fig. 2 shows the plots of $a_r$ and $b_{r+1}$ on $q$ for the most low-lying five bands [7]. As one can see, when $q = 11.2$ ($E \sim 28$ MeV), the fourth CP energy band can not be considered as a discrete "level", while the fifth CP energy band should remain a zone of the continuous spectrum.

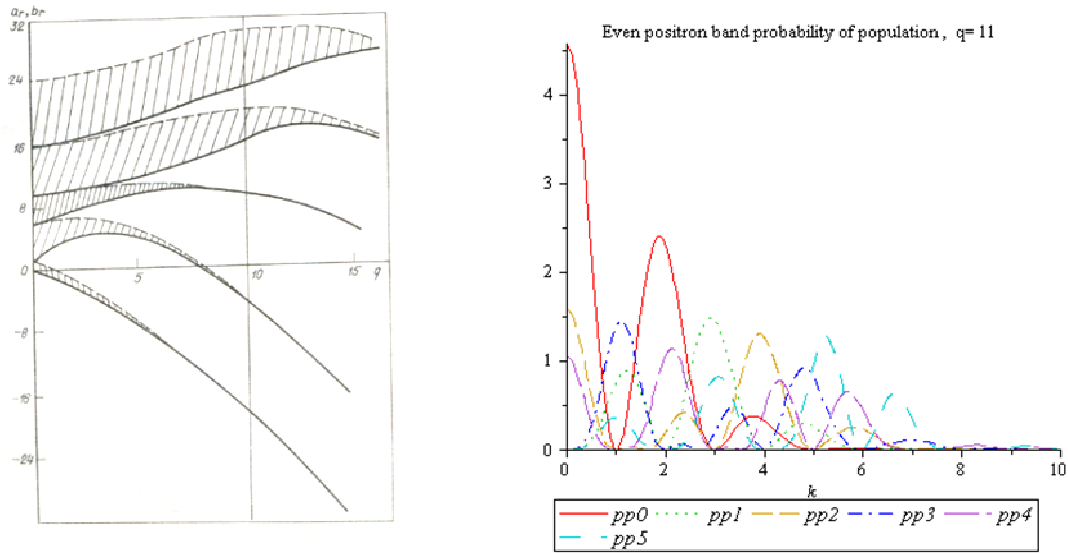

Fig.2. *(a) Energy bands (levels) of transverse motion of a fast oriented positrons for 28 MeV in planar channeling along the plane (110) in single crystal Si in the approximation of a sine crystal potential (with vertical line band spectrum for the energy of the particle $E \sim 25$ MeV is marked, i.e. $q_0 = 10$): – – – – $a_r$ - upper border of bands, _____ $b_r$ - lower border of bands, $q$ - quasi-momentum ($a_r, b_r, q$ in dimensionless units). (b) Population of energy zones for zero entrance angle of positron with respect to the plane (110).*

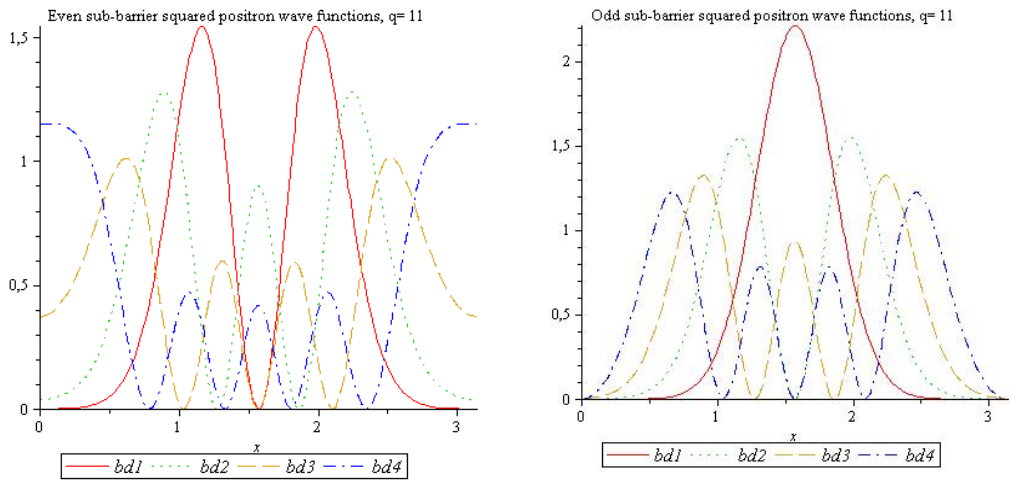

Fig.3. *The numerical simulation of the averaged wave function square module for different zones. Averaged square modules of odd and even wave functions for 28 MeV positrons in planar channeling along the plane (110) in single crystal Si for different energy bands: (1) – the 1-st deepest under-barrier level; (2) – level 4 in the middle of the channel; (3) – the highest under-barrier level; (4) – the first over-barrier band.*

From another hand, the particles having low transverse energies are moving between the atomic planes, where the electron density is low and electrons are mostly covalent, having small binding energy, which is not enough to open the possibility of one photon annihilation.

Thus the orientational dependence of the single photon annihilation of relativistic positrons in single crystal would look like depicted on Fig.5 with the account to the relativistic positron matrix elements behavior (Fig.4):

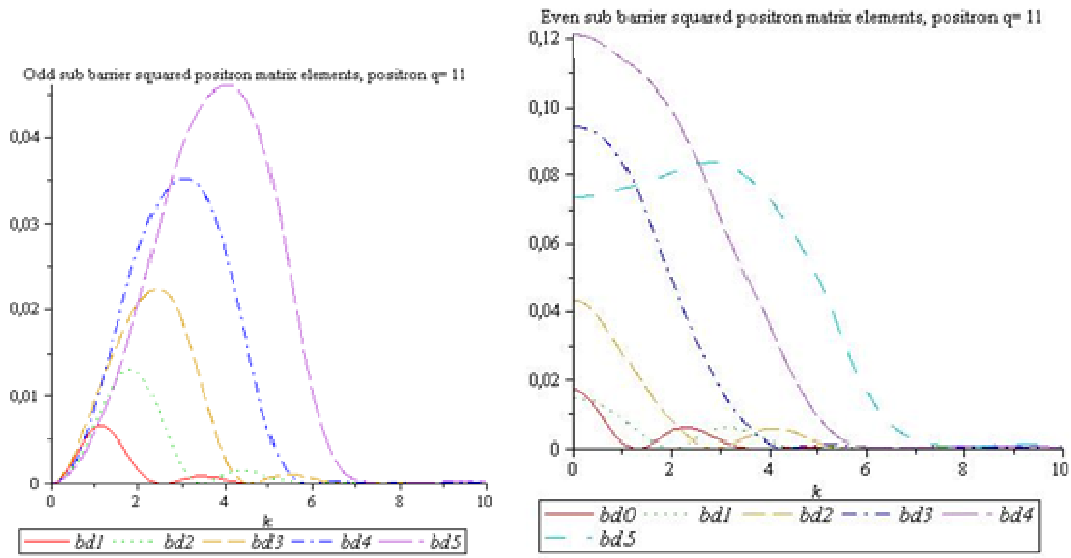

Fig.4. *Orientation dependence of the squared positron even and odd matrix elements on the positron entrance angle θ measured in inverse crystal lattice vectors.*

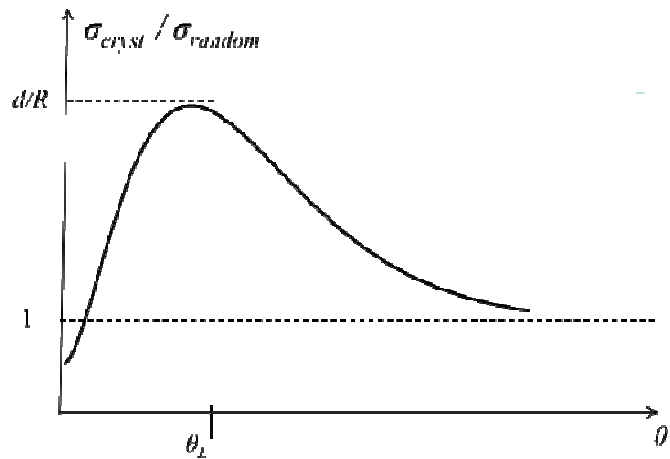

Fig.5. *Orientation dependence of single-photon annihilation at the positron entrance angle θ.*

### 4. Conclusions

As a result of the quantum-mechanical analysis we found that the dependence of the single-photon annihilation of relativistic positrons is a complicated function of the angle of incidence of the positron with respect to the crystallographic planes of the crystal. If the angle of incidence of the positron is smaller than the Lindhard angle, the probability of the single-photon annihilation is depressed due to the low density of electrons inthe channel between crystal planes. At angles of incidence close to the Lindhard angle we see the increase in the probability of single-photon annihilation due to hovering of most of positrons over crystal planes with high concentration of atomic electrons, as well as due to the behaviour of the quantum transition matrix elementsof the positron in the close to the barrier states. Orientation behaviour of the single-photon annihilation of positrons in a crystal is evidently pronounced and can easily be observed in the experiment.